# DATA-CONSISTENT NON-CARTESIAN DEEP SUBSPACE LEARNING FOR EFFICIENT DYNAMIC MR IMAGE RECONSTRUCTION

*Zihao Chen[1,2], Yuhua Chen[1,2], Yibin Xie[1], Debiao Li[1,2], and Anthony G. Christodoulou[1,2]*

[1]Biomedical Imaging Research Institute, Cedars-Sinai Medical Center, Los Angeles, USA
[2]Department of Bioengineering, UCLA, Los Angeles, USA

## ABSTRACT

Non-Cartesian sampling with subspace-constrained image reconstruction is a popular approach to dynamic MRI, but slow iterative reconstruction limits its clinical application. Data-consistent (DC) deep learning can accelerate reconstruction with good image quality, but has not been formulated for non-Cartesian subspace imaging. In this study, we propose a DC non-Cartesian deep subspace learning framework for fast, accurate dynamic MR image reconstruction. Four novel DC formulations are developed and evaluated: two gradient decent approaches, a directly solved approach, and a conjugate gradient approach. We applied a U-Net model with and without DC layers to reconstruct T1-weighted images for cardiac MR Multitasking (an advanced multidimensional imaging method), comparing our results to the iteratively reconstructed reference. Experimental results show that the proposed framework significantly improves reconstruction accuracy over the U-Net model without DC, while significantly accelerating the reconstruction over conventional iterative reconstruction.

*Index Terms*— MRI reconstruction, Deep learning, Non-Cartesian, Subspace, Dynamic MRI

## 1. INTRODUCTION

Dynamic magnetic resonance imaging (MRI) can be used to evaluate organ motion, tissue dynamic contrast enhancement (DCE), and nuclear magnetic resonance (NMR) relaxations. Dynamic imaging is therefore vital for applications such as cardiac and cancer imaging, as well as for relaxometry in any organ. The slow speed of MRI often results in high undersampling of the acquired k-t space signals, motivating constrained subspace/low-rank [1] and/or compressed sensing [2] reconstruction alongside acquisition schemes such as randomized and/or non-Cartesian sampling. The specific combination of non-Cartesian acquisition and subspace reconstruction is central to many promising imaging frameworks such as MR fingerprinting [3-5], MR Multitasking [6], Extreme MRI [7], and GRASP-Pro [8, 9].

Subspace methods alleviate many of the computational challenges of dynamic imaging by efficiently modeling, reconstructing, and storing dynamic images in a low-dimensional subspace rather than in the time domain. However, the use of non-Cartesian sampling counteracts this computational efficiency: non-Cartesian Fourier encoding operators such as the nonuniform fast Fourier transform (NUFFT) [10] have no direct inverse, so non-Cartesian subspace imaging is still largely performed via iterative reconstruction. The resulting long reconstruction time limits practical adoption of these imaging frameworks.

In recent years, deep learning reconstruction methods have shown great advantage over iterative methods in reducing reconstruction time [11-13]. Data-consistency (DC) layers, or unrolled deep learning, utilize the acquired data to adjust the network's output and have been shown to improve image quality and generalizability over purely training-data-driven deep-learning reconstruction [14-20]. The general DC problem has previously been formulated using gradient descent (GD) [16, 17, 19], density-compensated gradient descent (DGD) [20], or conjugate gradient (CG) [15, 18] approaches—and in the case of Cartesian encoding, as a coil-by-coil direct inversion layer [14, 15].

In the realm of subspace image reconstruction, Chen et al. have formulated non-Cartesian deep subspace learning without DC layers [21], and Sandino et al. [22, 23] have formulated DC deep subspace learning reconstruction for Cartesian trajectories using GD-DC or CG-DC layers. However, data-consistent non-Cartesian deep subspace learning remains a challenge: current non-Cartesian DC MRI strategies for dynamic imaging are formulated in k-t space [16, 18-20] rather than in the more computationally efficient subspace; GD based DC layers are inefficient due to ill-conditioning of the non-Cartesian reconstruction problem [24]; and iterative CG-DC layers offer slower reconstruction.

In this study, we developed and evaluated four novel formulations of data-consistent non-Cartesian deep subspace learning image reconstruction: GD-DC, preconditioned gradient descent (PGD-DC), directly-solved data consistency (DS-DC), and CG-DC. Inspired by Toeplitz encoding models [25], the DS-DC approach relies on an invertible block-Toeplitz model of the combined forward/adjoint encoding operator, allowing a coil-wise closed-form solution of the non-Cartesian DC equation. We compared these DC layers to each other and to non-DC deep learning in the context of cardiac MR Multitasking [26], using a U-Net [27] model as the network module.

## 2. THEORY

### 2.1. Non-Cartesian subspace image reconstruction

In subspace reconstruction for non-Cartesian dynamic MRI, a dynamic image represented as a matrix $X \in \mathbb{C}^{N_x N_y \times T}$ is decomposed into a spatial factor $U \in \mathbb{C}^{N_x N_y \times L}$ and a temporal factor $\Phi \in \mathbb{C}^{L \times T}$ according to $X = U\Phi$. When the rows of $\Phi$ constitute an orthonormal basis, $U$ can be interpreted as coordinates within the $L$-dimensional subspace spanned by the rows of $\Phi$. A suitable $\Phi$ can often be quickly extracted from a subset of acquired data $b$ via PCA or SVD [1], or calculated *a priori* [28], depending on the application. The most time-consuming step of reconstruction is therefore to estimate $U$, typically by solving:

$$\widehat{U} = \arg\min_U \|b - A_\Phi(U)\|_2^2 + \lambda R(U) \quad (1)$$
$$A_\Phi(U) = \Omega([F_{\text{NU}}SU]\Phi) \quad (2)$$

Here $F_{\text{NU}}$ is the NUFFT, $S$ applies sensitivity maps, $\Omega$ is the k-t space undersampling operator and $\lambda R(U)$ is a regularization term, often a sparse regularizer in order to leverage compressed sensing [29]. Conventionally, this minimization problem is solved by iterative methods such as alternating direction method of multipliers (ADMM) or the fast iterative soft-thresholding algorithm (FISTA).

### 2.2. Non-Cartesian deep subspace learning

Deep learning frameworks instead use a feedforward neural network to reconstruct $U$. This can be done, for example, by passing an initial guess $U_0$ through a convolutional neural network (CNN):

$$U_{\text{cnn}} = CNN(U_0), \quad (3)$$

where the network has been trained to produce an output $U_{\text{cnn}}$ that resembles the solution to Eq. (1) [21].

This network output $U_{\text{cnn}}$ can further pass through a data consistency (DC) layer that improves data fidelity by re-incorporating the measured k-space into the network's reconstruction. We formulate the DC problem similarly to Eq. (1), replacing $R(U)$ with $\|U - U_{\text{cnn}}\|_2^2$ in order to produce a CNN-regularized reconstruction:

$$\widehat{U} = \arg\min_U \|b - A_\Phi(U)\|_2^2 + \lambda \|U - U_{\text{cnn}}\|_2^2. \quad (4)$$

The solution can be expressed as:

$$\widehat{U} = (A_\Phi^* A_\Phi + \lambda I)^{-1}(A_\Phi^* b + \lambda U_{\text{cnn}}), \quad (5)$$

where the operator $A_\Phi^*$ is the conjugate transpose of $A_\Phi$.

However, Eq. (5) is difficult to solve analytically for non-Cartesian MRI. Time-consuming CG iterations could be used to solve Eq. (5), but would offset the reconstruction time advantages of using deep learning.

### 2.3. Non-Cartesian subspace DC layer with gradient descent methods

To avoid inverting $A_\Phi^* A_\Phi + \lambda I$, we can formulate the DC network in a gradient descent manner:

$$U_{\text{dc}} = U_{\text{cnn}} - \alpha[A_\Phi^* A_\Phi(U_{\text{cnn}}) - A_\Phi^*(b)]. \quad (6)$$

Eq. (6) subtracts the gradient of the data-fidelity term in Eq. (4) from the CNN output $U_{\text{cnn}}$ to improve data fidelity. $A_\Phi^* A_\Phi(U)$ is calculated according to:

$$A_\Phi^* A_\Phi(U) = S^H F_{\text{NU}}^H [\Omega^* \Omega(F_{\text{NU}} S U \Phi) \Phi^H]. \quad (7)$$

A subspace kernel method can efficiently compute Eq. (7) by calculating $F_{\text{NU}} S U$, right-multiplying $L \times L$ subsets of $\Omega^* \Omega(\cdot \Phi) \Phi^H$ for each unique k-space trajectory, and finally applying $S^H F_{\text{NU}}^H$ [29]. This keeps all calculations within the $L$-dimensional subspace and avoids the larger memory usage of the time domain.

It was established in previous work that the non-Cartesian reconstruction problem is ill-conditioned due to nonuniform density, and that adding a preconditioner can accelerate convergence in gradient descent [24]. As such, we also formulate preconditioned gradient descent for non-Cartesian subspace DC as:

$$U_{\text{dc}} = U_{\text{cnn}} - \alpha S^\dagger P[E_\Phi^* E_\Phi(SU_{\text{cnn}}) - E_\Phi^*(b)] \quad (8)$$
$$E_\Phi(Y) = \Omega([F_{\text{NU}}Y]\Phi) \quad (9)$$

Here $E_\Phi$ is the coil-wise encoding matrix, and $P$ is a preconditioner approximating the pseudoinverse of $F_{\text{NU}}^H F_{\text{NU}}$ by compensating its nonuniform weighting in k-space.

### 2.4. Non-Cartesian subspace directly-solved DC layer with inverse Block Toeplitz method

Inspired by the Toeplitz method that can significantly accelerate the calculation of $E^* E(x) = F_{\text{NU}}^H [\Omega^* \Omega(F_{\text{NU}} x)]$ for static images [25], here we formulate a block-Toeplitz model for $E_\Phi^* E_\Phi$ which can be analytically inverted, opening the door to a direct solution to non-Cartesian subspace DC problem. The static Toeplitz method models $E^* E$ as a linear shift-invariant system that performs a convolution and can therefore be represented by a Toeplitz matrix. This Toeplitz matrix can be represented as $E^* E = Z^H F^{-1} Q F Z$, where $F$ is the Cartesian FFT, $Z$ zero-pads to twice the image size in each spatial dimension to accommodate circular convolution boundaries, and where $Q$ is a diagonal matrix that performs a k-space multiplication derived from the FFT of the point spread function (PSF).

The block-Toeplitz model for $E_\Phi^* E_\Phi$ combines the Toeplitz model of non-Cartesian forward/adjoint encoding and the $L \times L$ subspace kernel concept to express $E_\Phi^* E_\Phi$ as $L \times L$ block-Toeplitz, with the $(i,j)$-th block of $E_\Phi^* E_\Phi$ taking the form:

$$[E_\Phi^* E_\Phi]_{i,j} = Z^H F^{-1} Q^{(i,j)} F Z, \quad (10)$$

where $Q^{(i,j)}$ applies the k-space filter for that block. Equivalently, we can say that $E_\Phi^* E_\Phi(Y) = Z^H F^{-1} W(FZY)$, where $W(\cdot)$ right-multiplies an $L \times L$ kernel $W^{(n)}$ with elements $w_{ij}^{(n)} = q_{nn}^{(i,j)}$ at the $n$th of $2N_x \cdot 2N_y$ k-space locations.

Then we can consider a coil-wise DC equation

$$\widehat{Y} = \arg\min_Y \|E_\Phi(Y) - b\|_2^2 + \lambda \|Y - SU_{\text{cnn}}\|_2^2 \quad (11)$$

which has the solution

$$\widehat{Y} = (E_\Phi^* E_\Phi + \lambda I)^{-1}(E_\Phi^*(b) + \lambda SU_{\text{cnn}}). \quad (12)$$

The operator $E_\Phi^* E_\Phi + \lambda I$ can be directly inverted by regularized inversion of each $L \times L$ kernel $W^{(n)}$, i.e., as:

$$\hat{Y} = Z^H F^{-1}(W + \lambda I)^{-1}(FZ[E_\Phi^*(b) + \lambda S U_{cnn}]) \quad (13)$$

Where the function $(W + \lambda I)^{-1}(\cdot)$ right-multiplies an $L \times L$ kernel $(W^{(n)} + \lambda I)^{-1}$ at the $n$th k-space location.

Then, the data-consistent spatial factor $U_{dc}$ can be calculated by complex coil combination ($U_{dc} = S^\dagger \hat{Y}$). This method directly solves the non-Cartesian DC problem.

## 3. EXPERIMENTS

In this work, we evaluated three DC options described in the Theory section above: vanilla GD-DC based on Eq. (6), PGD-DC based on Eq. (8), and DS-DC based on the block-Toeplitz inversion in Eq. (13). We further evaluated a CG-DC layer based on Eq. (5) with 5 CG iterations.

### 3.1. Datasets

All data were dynamic MR cardiac images acquired with a T1 MR Multitasking protocol [26] on three different 3T MRI scanners (MAGNETOM Verio, MAGNETOM Vida, and Biograph mMR; Siemens Healthcare, Erlangen, Germany) at the same center. The k-t space data were acquired with a continuous IR-FLASH sequence and golden-angle radial trajectories. Label images were reconstructed iteratively as described in [26] with $L = 32$, during which the temporal factors were generated from dictionaries and auxiliary data [6, 26]. This produced a multidimensional array of images at each combination of $c = 20$ cardiac phases, $r = 6$ respiratory phases and $\tau = 344$ T1-recovery timepoints. Thus, there are $T = 20 \times 6 \times 344 = 41{,}280$ temporal frames for each dynamic image. The image matrix size for each frame is $320 \times 320$, corresponding to a FOV of $540 \times 540$ mm$^2$ (twice the prescribed FOV of $270 \times 270$ mm$^2$).

For deep subspace reconstruction, we directly fed spatial factors $U$ rather than images $X$ into the network. We concatenated the real and imaginary parts of $U_0$'s and of labeled $U$'s as separate channels to preserve complex values, for network input and output sizes of $320 \times 320 \times 64$. In total there were 120 dynamic image sets, in which 96 sets were used for training, 12 sets for validation and 12 sets for testing.

### 3.2. Evaluation metrics

Since the networks were trained on spatial factors $U$ but the final dynamic images are calculated as $X = U\Phi$, we did comparisons for both $U$ and for reconstructed dynamic images. The reference images for our comparisons were iteratively reconstructed using wavelet sparsity regularization and 20 iterations of an ADMM algorithm.

For spatial factor/subspace coordinates $U$, we used normalized root-mean-square error (NRMSE) to evaluate the networks. For dynamic images, peak signal-to-noise ratio (PSNR), structural similarity index (SSIM) and NRMSE were calculated from the reconstructed image sequences for the whole cardiac cycle (20 frames) at the end-expiration (EE) respiratory phase, and for inversion times corresponding to bright-blood and dark-blood contrast weighting (i.e., the two most clinically important qualitative image contrasts).

### 3.3. Experimental setup

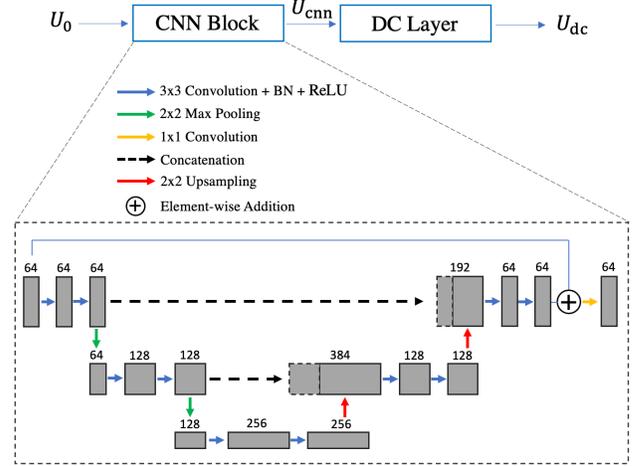

**Fig. 1.** Network architecture in proposed methods

All the proposed DC networks consist of a CNN Block (U-Net) and a DC layer (Fig. 1), which were implemented in TensorFlow. The input $U_0$ was obtained by zero-filled regridding: $U_0 = S^\dagger F_{NU}^H D\Omega^*(b)\Phi^H$, where $D$ applies a density compensation filter.

To compare the performance of different DC layers, we pretrained the U-Net block without DC layer for 200 epochs (5 hours) on one Nvidia Titan RTX GPU with 24GB of RAM, and directly added different DC layers at the end of the U-Net block without further training. Adam optimizer and mean squared error (MSE) loss were used in training. The step size $\alpha$ in GD-DC and PGD-DC, as well as the regularization coefficient $\lambda$ in DS-DC and CG-DC, were determined by choosing the best values for the validation set. The preconditioner in PGD-DC was a Cartesian k-space ramp which adjusted for radial k-space density. A single DC layer was used for each network to minimize reconstruction time.

In the comparison, four proposed DC networks (GD-DC, PGD-DC, DS-DC and CG-DC) and the pretrained U-Net without DC layer were compared with iterative reconstructed reference images among the testing set.

## 4. RESULTS

The spatial factor reconstruction time is shown in Table 1. GPU memory use of each DC layer was 1.6 GB. All the single-step DC models accelerated the reconstruction time by more than 50x compared to 180 sec iterative reconstruction.

For quantitative comparison among the testing set, Table 1 shows the NRMSE of different models for the spatial factor $U$ directly output by the network; Table 2 shows the PSNR,

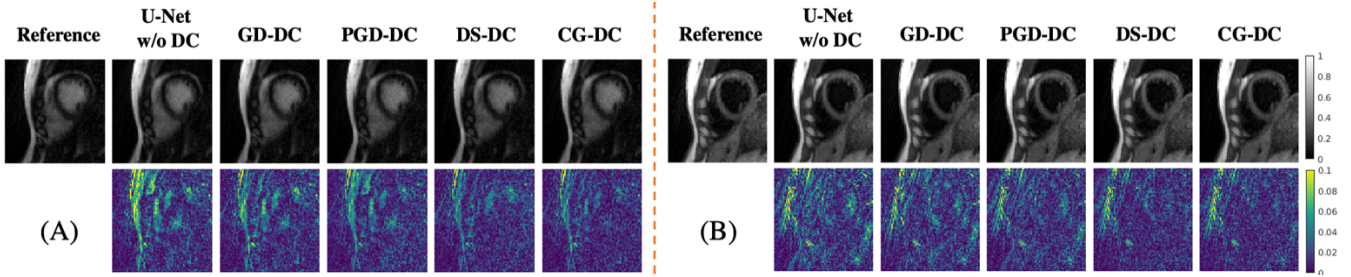

**Fig. 2.** Example T1-weighted images from iterative reconstruction and different networks with corresponding error maps. (A): Bright blood contrast images; (B): dark blood contrast images. Top row: examples images; bottom row: corresponding error maps.

SSIM and NRMSE of different models for reconstructed dynamic images for bright blood and dark blood contrast weightings. Since each dynamic image contains 20 cardiac frames, we have 12×20=240 testing images for each contrast in Table 2. CG-DC had the best quantitative metrics ($p<0.001$), but its computation time was >6x that of the single-step DC layers. Among the single-step DC layers, DS-DC performed best ($p<0.001$), followed by PGD-DC, GD-DC and U-Net w/o DC, for both the spatial factor and dynamic images.

**Table 1.** NRMSE over spatial factor $U$ among the testing set. Values in brackets are standard deviations. Best results among single-step DC models are bolded. Best results among all models are italic.

|  | No DC | Single-step DC | | | Iterative DC |
|---|---|---|---|---|---|
| **Model** | **Input $U_0$** | **U-Net w/o DC** | **GD-DC** | **PGD-DC** | **DS-DC** | **CG-DC** |
| NRMSE | 0.279 (0.096) | 0.169 (0.052) | 0.156 (0.053) | 0.152 (0.050) | **0.135 (0.041)** | *0.117 (0.036)* |
| Inference Time | N/A | *1.7s* | **3.5s** | **3.5s** | **3.5s** | 22s |

**Table 2.** Quantitative metrics over dynamic images among the testing set. Values in brackets are standard deviations. Best results among single-step DC models are bolded. Best results among all models are italic.

|  | | PSNR | | SSIM | | NRMSE | |
|---|---|---|---|---|---|---|---|
|  | Model | Bright Blood | Dark Blood | Bright Blood | Dark Blood | Bright Blood | Dark Blood |
| **No DC** | U-Net w/o DC | 33.67 (2.95) | 36.89 (2.60) | 0.854 (0.057) | 0.915 (0.024) | 0.174 (0.054) | 0.108 (0.032) |
| **Single-step DC** | GD-DC | 34.14 (3.03) | 37.44 (2.69) | 0.873 (0.053) | 0.926 (0.024) | 0.165 (0.053) | 0.102 (0.032) |
|  | PGD-DC | 34.31 (3.00) | 37.76 (2.44) | 0.873 (0.051) | 0.927 (0.022) | 0.163 (0.058) | 0.098 (0.028) |
|  | DS-DC | **35.37 (2.87)** | **38.68 (2.41)** | **0.888 (0.040)** | **0.933 (0.017)** | **0.144 (0.048)** | **0.088 (0.024)** |
| **Iterative DC** | CG-DC | *35.73 (2.83)* | *39.25 (2.29)* | *0.897 (0.040)* | *0.945 (0.014)* | *0.138 (0.044)* | *0.082 (0.022)* |

Fig. 2 shows an example testing case of the T1-weighted images at EE respiratory phase and end-diastolic cardiac phase for bright blood and dark blood contrasts. The visual comparison is consistent with the quantitative results: CG-DC and DS-DC have the smallest errors, followed by PGD-DC, GD-DC and U-Net w/o DC. The error maps of CG-DC and DS-DC have fewer structural features than those of other models, implying that CG-DC and DS-DC provided less systematic error.

## 5. DISCUSSION & CONCLUSIONS

In this study, we developed a DC non-Cartesian deep subspace learning framework to accelerate dynamic MR reconstruction and proposed four DC approaches: GD-DC, PGD-DC, DS-DC and CG-DC. All the deep learning models except CG-DC accelerated reconstruction by more than 50x over iterative reconstruction. All DC models outperformed the naïve U-Net w/o DC in quantitative comparisons. CG-DC had the least error but longest inference time (22 s), while DS-DC provided the best accuracy amongst the fast (3.5 s) single-step DC layers. CG-DC may be desirable when imaging a single slice with one inference, whereas DS-DC may be more attractive when imaging multiple slices.

All the subspace DC formulations substantially reduced the memory required for large-scale dynamic image reconstruction compared to direct implementation of previously proposed DC layers in the time domain. A DC layer in k-t space here would have required operating on the 8,200 readout time points rather than the $L = 32$ entries in $U$, requiring 410 GB of memory instead of our 1.6 GB.

Although our DC layers were applied in subspaces here, the proposed inverse block-Toeplitz DS-DC can be readily adapted to time-domain or static imaging to improve the efficiency of general non-Cartesian deep learning.

In this work, we chose U-Net as our CNN block for simplicity. The proposed DC layers can be easily added to other advanced CNN blocks to further improve their reconstruction quality. This study also only evaluated a single pre-trained CNN+DC block, but multiple CNN+DC blocks and end-to-end training may offer even further improvement.

In conclusion, the proposed DC deep subspace learning framework significantly improves reconstruction accuracy over the plain U-Net model, while significantly accelerating reconstruction over conventional iterative algorithms. Clinical studies are needed to evaluate the diagnostic accuracy and clinical value of the proposed deep learning reconstruction model.

## 6. COMPLIANCE WITH ETHICAL STANDARDS

Informed consent was obtained for all human subjects in accordance with an institutional review board protocol at Cedars-Sinai Medical Center.

## 7. ACKNOWLEDGMENTS

This work was supported by NIH R01 EB028146.